\documentclass[useAMS,usegraphicx]{mn2e}
\usepackage{times}

\def\fQPO{f_{\rm QPO}}
\def\Tbb{T_{\rm bb}}
\def\kTbb{k T_{\rm bb}}
\def\kT{k T_{\rm e}}
\def\Ka{K$\alpha$\ }
\def\Rin{R_{\rm in}}
\def\lheat{l_{\rm heat}}
\def\lsoft{l_{\rm soft}}
\def\Or{R}
\def\chir{\chi^2_{\nu}}

\title[QPO X-ray spectra]
   {Spectral and Fourier analyses of X-ray  quasi-periodic oscillations
     in accreting black holes}

\author[M.A. Sobolewska and  P.T.\.Zycki]
 {M. A. Sobolewska\thanks{E-mail:malsob@camk.edu.pl}\thanks{Present address:
  University of Durham, Physics Department, South Road, Durham DH1 3LE, UK}
   and P. T. \.Zycki\thanks{E-mail:ptz@camk.edu.pl}\\
Nicolaus Copernicus Astronomical Center, Bartycka 18, 00-716 Warsaw, Poland}

\begin{document}

\date{9 May 2006}

\voffset -1.5 cm

\pagerange{\pageref{firstpage}--\pageref{lastpage}} \pubyear{2005}

\maketitle

\label{firstpage}

\begin{abstract}

We study energy dependencies of quasi-periodic
oscillations from a number of black hole 
X-ray binaries. The selected sources were observed by {\it 
RXTE\/} at time periods close to state transitions and showed QPO in the 
1--10 Hz range. We have constructed QPO root-mean-square energy spectra, 
which provide information about underlying physical process leading to QPO 
generation. These spectra show an interesting  
anti-correlation with the time averaged spectra. The QPO r.m.s.\ spectra 
are harder than the time averaged spectra when the latter are soft, while 
they are softer than the time averaged spectra when the latter are hard. 
We then discuss these 
observational results in the context of simple spectral variability 
models. Hard QPO spectra can be produced by quasi-periodic modulations
of the heating rate of the Comptonizing plasma, while soft QPO spectra
result from modulations of the cooling rate by soft photons.

\end{abstract}


\section{Introduction}


A common feature in X-ray power spectra from many X-ray binaries (XRB) are 
quasi-periodic oscillations (QPO). Many  
``flavours'' of QPO have been identified, with some correlations between 
their frequencies (Psaltis, Belloni \& van der Klis 1999). For example, 
low frequency ($f$ of a few Hz) QPO in black hole XRB seem to appear 
preferentially when the state of the source changes (Rutledge et al.\ 
1999). They are mostly observed in luminous states, high or very high 
state, when the energy spectrum is rather complex, showing both disc 
thermal component and high energy Comptonized component (see QPO observations
review in Wijnands 2001). Models of QPO so far concentrated on identifying the 
frequencies of oscillations, with very little if any attention paid to the 
fact that it is the hard X--rays which are modulated (see QPO models review in 
Psaltis 2001). The results of Gilfanov, Revnivtsev \& Molkov (2003) and 
Revnivtsev \& Gilfanov  (2005) are perhaps the only ones where the energy 
spectra of QPO are discussed and interpreted. These authors applied
Fourier frequency resolved spectroscopy (Revnivtsev, Gilfanov \&
Churazov 1999, 2001) to show in particular that the kilo-Hz
QPO spectra in neutron star XRB 
are the same as spectra at other Fourier frequencies.  This can be 
interpreted as common spatial origin of the X-rays showing the usual broad 
band noise variability and the more coherent quasi-periodic variability.
The spectral shape of these components is consistent with Comptonized
emission from a boundary layer.

Fourier spectroscopy techniques are particularly suitable for analysis of 
QPO. However, the approach of Revnivtsev et al.\ (1999) requires
integrating the power density spectra over a narrow frequency interval
around a chosen frequency resulting in contamination of the Fourier
frequency resolved spectrum of the QPO by a contribution from the broad
band variability component. Thus, in this paper we construct a
root-mean-square spectrum, where each point is the r.m.s.\  variability 
integrated over the component (e.g.\ a Lorentzian) describing the QPO in 
the broad band power spectrum. Such energy dependences of QPO amplitude 
in XRB has been investigated before (e.g.\ Belloni et al.\ 1997; 
Cui et al.\ 1999; Rao et al.\ 2000; Tomsick \& Kaaret 2001; Rodriguez et
al.\ 2004; Casella et al.\ 2004; see review in van der Klis 2006). 
The main conclusion from these
studies is that the QPO amplitude increases with energy up to at least
$\sim 20$ keV. This suggests immediately that the oscillations are connected
with the high energy spectral component, rather than with the disk
component.

\begin{table*}
\caption{Log of RXTE observations}
\centering
\begin{tabular}{l c c c c}
\hline
Object & MJD$^a$ & yymmdd & Observation ID & Spectral state\\
\hline
XTE~J1859+226 (I) & 51463.8 & 991012 & 40124-01-05-00 & soft state at the luminosity peak$^1$\\ 
XTE~J1859+226 (II) & 51465.9 & 991014 & 40124-01-10-00 & ''\\
XTE~J1859+226 (III) & 51472.5 & 991021 & 40124-01-21-00 & ''\\ 
XTE~J1859+226 (IV) & 51474.8 & 991023 & 40124-01-25-00 & ''\\
\hline
GRS~1915+105 (I) & 50422.0 & 961204 & 20402-01-05-00 & C state (class $\chi$)$^2$\\
GRS~1915+105 (II)  & 52768.6 & 030509 & 80127-03-01-00 & '' $^3$\\
GRS~1915+105 (III) & 52731.5 & 030402 & 80127-01-03-00 & '' \\
\hline
XTE~J1550-564 (I) & 51068.3 & 980912 & 30188-06-05-00 & soft state$^4$\\
XTE~J1550-564 (II) & 51245.4 & 990308 & 40401-01-53-00 & ,,anomalous'' state$^5$\\
XTE~J1550-564 (III) & 51248.1 & 990311 & 40401-01-51-01 & ''\\
XTE~J1550-564 (IV) & 51250.7 & 990313 & 40401-01-57-00 & ''\\
XTE~J1550-564 (V) & 51683.8 & 000519 & 50135-01-03-00 & hard state$^6$\\
\hline
4U~1630-47 (I)   & 50853.1 & 980209 & 30178-01-01-00 & hard state$^7$\\ 
4U~1630-47 (II)  & 50853.7 & 980209 & 30178-01-02-00 & '' \\
\hline
\end{tabular}\\
$^a$ Start of observation, ${\rm MJD} = {\rm JD} - {\rm 2,400,000.5}$\\
References to papers with spectral/timing analysis:\\
$^1$ Casella et al.\ (2004)\\
$^2$ Sobolewska \& \.{Z}ycki (2003)\\
$^3$ Rodriguez et al.\ (2004)\\
$^4$ Sobczak et al. (2000)\\
$^5$ Remillard et al.\ (2002), Kubota \& Done (2004)\\
$^6$ Tomsick et al. (2001), Rodriguez et al.\ (2003)\\
$^7$ Dieters et al.\ (2000)\\
\label{tab:data}
\end{table*}

The significance of the r.m.s.\ energy spectrum is that it
describes the energy spectrum of the process responsible for a given
variability component (e.g.\ a QPO), if the variability is generated 
by fluctuactions of the normalization of a separate spectral component
(Gilfanov et al.\ 2003; \.{Z}ycki \& Sobolewska 2005, hereafter Paper I). 
If the QPO is produced by an additional modulation of one or more physical 
parameters determining the spectrum, its r.m.s.\ spectrum has less direct 
interpretation, but still it contains the signatures of the modulation 
(Paper I).
Energy dependencies of r.m.s.\ spectra were investigated by many
authors with a goal of determining how the spectral components vary
with respect to each other. Recently, Gierli\'{n}ski \&
Zdziarski (2005) investigated the r.m.s.\ energy spectra of a number
of XRBs in various spectral stated and interpreted them in terms of
variations of plasma heating/cooling rate in a scenario where the time
average energy spectrum consists of a disc blackbody and a hybrid
(thermal/non-thermal) Comptonization. However, their r.m.s.\ spectra
were averaged over all variability components (Fourier frequency). 

In this paper we perform a systematic study of energy dependencies of
low-frequency QPO in black hole XRB in different spectral states.  
We analyse {\it RXTE\/} data from a 
number of sources observed at time periods close to or during state 
transitions, when the sources showed QPO in 1--10 Hz range. 
We selected data with good 
energy and time resolutions, enabling construction of power density 
spectra (PDS) in at least 12 energy channels covering the 3--40 keV range. 
We then constructed the r.m.s.\ spectra of 
the QPO and compared them with the mean energy spectra. Finally, we 
compared the results of data analysis with the predictions of theoretical 
models. The models themselves are described in detail in Paper I.

\begin{table*}
\caption{Fits to time averaged spectra}
\centering
\begin{tabular}{l c c c c c c c c}
\hline
  &   \multicolumn{3}{c}{diskbb/thComp 1} & \multicolumn{4}{c}{thComp 2} & \\
Dataset & $\kTbb$ (keV)  & $\Gamma$ & $\kT$ (keV)  &
 $\Gamma$ & $\kT$ (keV)  & $\Or$ & $\xi$ & $\chi^2/{\rm dof}$ \\ 
\hline
XTE~J1859+226 (I)$^a$  & $0.70^{+0.09}_{-0.16}$  & --& -- & 
     $2.19\pm 0.08$  & $44^{+100}_{-18}$ &
     $0.19^{+0.13}_{-0.10}$ &  $185^{+4700}_{-110}$  & 73.9/69 \\        
XTE~J1859+226 (II)$^a$  & $0.81 \pm 0.04$ & -- & -- & 
   $2.38 \pm 0.03$ & 150(f) & 
   $0.15^{+0.07}_{-0.03}$  & $(4.4^{+9.6}_{-3.9})\times 10^3$  &  61.2/67 \\
XTE~J1859+226 (III)$^a$ & $0.82\pm 0.03$  & -- & -- &
   $2.37\pm 0.02$ & 150(f) & 
   $0.16 \pm 0.03$  &  $(2.5^{+5.5}_{-1.6})\times 10^3$ & 74.6/70   \\
XTE~J1859+226 (IV)$^a$  & $0.83\pm 0.03$ & -- & -- & $2.34\pm 0.02$ & 150(f) & 
   $0.16^{+0.06}_{-0.02}$  & $(2.9^{+6.0}_{-2.0})\times 10^3$  & 80.5/70  \\
\hline       
XTE~J1550-564 (I)$^c$  & $0.64^{+0.06}_{-0.20}$ & $1.98 \pm 0.12$ & 
 $7.5^{+1.2}_{-0.9}$ &
  $2.23^{+0.10}_{-0.37}$  & 150 (f) & $0.43^{+3.2}_{-0.15}$ & 
       $(15^{+25}_{-10})\times 10^3$ &  58.2/70 \\
XTE~J1550-564 (II)$^c$  & $0.92\pm 0.02$ & $2.86^{+0.24}_{-0.19}$ & 
 $17^{+51}_{-12}$ &
  $1.88 \pm 0.20$  & 150(f) & $> 0.8$ & 
    $(2.1^{+4.0}_{-1.7}\times 10^3$ & 63.5/69 \\
XTE~J1550-564 (III)$^b$ & $0.66\pm 0.01$  & $4.51 \pm 0.17$  & 100(f) &
  $2.19 \pm 0.05$ & 150(f)  & $0.40^{+0.33}_{-0.08}$ & 
      $(4.4^{+15.6}_{-3.6}) \times 10^3$ & 51.2/73 \\
XTE~J1550-564 (IV)$^b$ & $0.47 \pm 0.03$ & $4.1 \pm 0.4$ & 100(f) &  
 $2.21\pm 0.04$ 
 & 150(f) & $0.29^{+0.34}_{-0.04}$ & $(1.2^{+7.8}_{-1.0})\times 10^3$ 
         & 74.5/74 \\
\hline
GRS~1915+105 (I)$^a$ & $0.58^{+0.13}_{-0.08}$ & -- & -- & $2.29 \pm 0.07$ &
 $40^{+29}_{-20}$ &
 $0.47^{+0.73}_{-0.20}$ & $(2.0^{+4.0}_{-1.7}) \times 10^4$ & 59.7/71 \\
GRS~1915+105 (II)$^c$ & $0.19^{+0.13}_{-0.19}$ & $2.21^{+0.07}_{-0.05}$ & 
  $5.1^{+0.4}_{-0.9}$ &
  $2.25^{+0.09}_{-0.32}$ & $> 25$ & $1.7^{+1.7}_{-1.1}$ 
  & $(7.3^{+10}_{-4.6})\times 10^4$ & 69.2/62 \\
GRS~1915+105 (III)$^c$ & $1.02^{+0.24}_{-0.17}$ & $2.18^{+0.16}_{-0.50}$ & 
 $4.1^{+0.4}_{-0.6}$ &
 $2.50^{+0.40}_{-0.25}$ & 150(f) & $> 0.25$ & 
$(3.5^{+3.7}_{-2.1} \times 10^4$ & 44.5/65 \\
\hline       
\hline       
4U~1630-47 (I)$^a$    & $1.06^{+0.18}_{-0.30}$ & -- & -- & 
   $1.98^{+0.02}_{-0.04}$  & 
   $> 45$ &   $0.10^{+0.06}_{-0.02}$  & 
   $(4.9^{+8.1}_{-4.5})\times 10^3$  & 80.9/73  \\
4U~1630-47 (II)$^a$   & $1.05^{+0.22}_{-0.18}$  & -- & -- & $1.98 \pm 0.04$ &
   $37^{+40}_{-15}$ & $0.07^{+0.10}_{-0.01}$  &
   $(1.8^{+6.0}_{-1.6})\times 10^3$ & 44.7/66   \\
XTE~J1550-564 (V)$^a$ & $0.88^{+0.11}_{-0.09}$ &  -- & -- & 
 $1.677^{+0.019}_{-0.010}$ & 
 $> 64$ &  $0.11 \pm 0.03$ & 
  $(1.7^{+5.0}_{-1.1})\times 10^3$ & 41.8/69 \\
\hline
\end{tabular}\\
$^a$ Model: diskbb + thComp \\
$^b$ Model: thComp + thComp \\
$^c$ Model: diskbb + thComp + thComp \\
(f) Parameter was fixed\\
\label{tab:specaver}
\end{table*}

\section{Data selection and description of time-averaged spectra}

\label{sec:datasel}

We made use of the {\it RXTE\/} archive data. We were particularly 
interested in observations during which a pronounced (rms $>$ 10\%) and 
coherent 1--10 Hz QPO were present in the power density spectrum. A basic 
criterion was good timing {\it and\/} spectral resolution of the data. The 
log of all observations is presented in Table~\ref{tab:data}.

We extracted Standard 2 PCA spectra from top and mid layer from all available
units. Standard selection criteria and dead time correction procedures were 
applied. The PCA background was estimated using the {\it pcabackest\/} package
(ver.\ 3.0), while the response matrices were made for each observation
with {\it pcarsp\/} ver.\ 10.1 tool. Systematic error of 0.5 per cent were
added to each bin of PCA spectra, to take into account possible
calibration uncertainties.
HEXTE data (Archive mode, Cluster 0) were extracted using the same selection
criteria as those for the PCA. Background was estimated using {\it hxtback\/}
tool.
For spectral modelling we used the PCA data in 3--20 keV range and HEXTE data 
20--200 keV range.

The spectral model consisted of a disc blackbody and thermal Comptonization 
components, modified
by photoelectric absorption.  For Comptonization we use the {\sc thComp} 
XSPEC model (Zdziarski, Johnson \& Magdziarz 1996). 
This is parametrized by the asymptotic spectral index, $\Gamma$, seed photon
temperature, $T_0$, and electron temperature, $k T_{\rm e}$. Disc blackbody
(maximum) temperature is denoted $\Tbb$. In most cases we assume 
$T_0 = \Tbb$ as can be expected, if it is the disc photons which are 
Comptonized, but we will also consider cases where the temperatures
are de-coupled.  The {\sc thComp} model computes also 
self-consistently the reprocessed component consisting of the Compton 
reflected continuum and the Fe fluorescent K$\alpha$ line (\.{Z}ycki, Done 
\& Smith 1999 and references therein). The main parameters here are 
the relative amplitude of the component, $\Or$, ionization parameter
of the reprocessor, $\xi$, and the inner disc radius used for computing
the smearing of spectral features due to relativistic effects, $\Rin$.

Results of the fits to the time averaged spectra are given in 
Table~\ref{tab:specaver} and described in detail in following Sections. All
uncertainties in the Table are 90\% confidence limits for one parameter of 
interest, i.e.\ $\Delta\chi^2=2.71$.
We allowed for a free normalization between the data from PCA and HEXTE. 

\subsection{XTE~J1859+226}

We study four observations of the X-ray nova XTE~J1859+226 at the 
luminosity peak during its 1999 outburst. Detailed timing analysis and 
identification of different types of QPO was presented in Casella et 
al.\ (2004). During the observations the source was in a very high state. 
The data are well described with a disc blackbody model with a temperature 
of $\sim 0.7$--0.8 keV and its thermal Comptonization on electrons with 
temperature fixed at $kT_{\rm e} = 150$ keV (lower temperature, 
$kT_{\rm e} \approx 40$ keV is required for the 
first observation). The Comptonized continuum has a photon index of 
$\Gamma = 2.1$-- 2.4. In all datasets a weak (amplitude 
$\Or = 0.1$--0.2), highly ionized (ionization parameter 
$\xi = 1200$--2700) reflection component is present. All 
but the first observation require an additional smearing, which we model
as relativistic effects.

\subsection{GRS~1915+105}

We analysed three observations of GRS 1915+105. 
This is a very peculiar source 
showing complicated variability patterns and complex spectra. It seems to 
be always in a high/soft state, because  of high accretion rate
(Sobolewska \& \.{Z}ycki 2003; Done, Wardzi\'{n}ski \& Gierli\'{n}ski 
2004). The spectral analysis of the three  characteristic 
spectral states of this source, introduced by Belloni et al.\ (2000),
 was presented in Sobolewska \& \.{Z}ycki (2003).

The spectrum from the first observation can be well modelled as a sum of disc 
blackbody and a thermal Comptonization component. 
In such a scenario, disc blackbody 
photons with a temperature of $\sim 0.6$ keV provide seed photons for 
Comptonization on electrons with $k T_{\rm e} \approx 40$ keV. The continuum 
is modified by reflection ($\Or \approx 0.5$) from a highly ionized medium 
($\xi \sim 10^{4}$), with relativistic smearing. An additional 
gaussian line at 6.4 keV is also needed to fit the residua.

The same model does not describe well the two remaining datasets. Large 
residua in soft energy band remain and the model cannot fit simultaneously 
the high energy tail (above 50 keV). We try a  two-Comptonization model and 
a model consisting of a disc blackbody and two Comptonizations. While the
three component models give fits of comparable or marginally better quality
($\chir=69.5/61$ vs.\ 72.2/62 and 44.1/65 vs.\ 52.0/66, 
for datasets II and III, respectively), the best fit values of 
the reflection amplitude are closer to 1 than for the two component models.
We thus use the three component models in our further studies.
Disc photons 
temperature is $kT_{\rm bb} = 0.2$--1 keV. Electron temperature in the
soft Comptonization component is 4--5 keV, while in the hard Comptonization
component it is 50--150 keV. Best fit reflection amplitudes are high, 
$\Or=1.3$--1.7,although with quite high uncertainties, 
and the reflection is highly ionized, $\xi \sim 10^{4}$.
Additional narrow gaussian line at 6.4 keV of EW$\approx 50$ eV is also
required.

\begin{table*}
\caption{Analysis of r.m.s. spectra: frequency ranges
for the QPO spectra and PCA configurations for high time resolution
spectral data}
\centering
\begin{tabular}{l c l l}
\hline
Dataset               & f$_{\rm QPO}$ (Hz)  &~~~binned mode         & ~~~event mode \\ 
\hline
XTE~J1829+226 (I)     & 2.2--3.5  & B\_8ms\_16A\_0\_35\_H & E\_16$\mu$s\_16B\_36\_1s  \\
XTE~J1829+226 (II)    & 2.2--3.5  & B\_4ms\_16A\_0\_35\_H & E\_16$\mu$s\_16B\_36\_1s \\
XTE~J1829+226 (III)   & 4.3--7.2  & B\_8ms\_16A\_0\_35\_H & E\_16$\mu$s\_16B\_36\_1s \\
XTE~J1829+226 (IV)    & 4.8--7.2  & B\_8ms\_16A\_0\_35\_H & E\_16$\mu$s\_16B\_36\_1s \\
\hline
GRS~1915+105 (I)$^a$  & 2.2--4.3  & B\_8ms\_16A\_0\_35\_H & E\_62$\mu$s\_32M\_36\_1s \\
GRS~1915+105 (II)     & 2.0--3.6  & B\_4ms\_16A\_0\_35\_H & E\_500$\mu$s\_64M\_36\_1s  \\
GRS~1915+105 (III)    & 1.5--2.5  & B\_4ms\_16A\_0\_35\_H & E\_500$\mu$s\_64M\_36\_1s\\
\hline
XTE~J1550-564 (I)$^b$ & 2.0--3.1  & B\_4ms\_8A\_0\_35\_H & E\_16$\mu$s\_16B\_36\_1s \\
XTE~J1550-564 (II)    & 4.9--7.8  & B\_4ms\_8A\_0\_35\_H & E\_16$\mu$s\_16B\_36\_1s \\
XTE~J1550-564 (III)   & 4.4--7.1  & B\_4ms\_8A\_0\_35\_H & E\_16$\mu$s\_16B\_36\_1s \\
XTE~J1550-564 (IV)    & 5.5--8.3  & B\_4ms\_8A\_0\_35\_H & E\_16$\mu$s\_16B\_36\_1s  \\
XTE~J1550-564 (V)     & 1.9--5.2  & B\_8ms\_16A\_0\_35\_H & E\_125$\mu$s\_64M\_0\_1s \\
\hline       
\end{tabular}\\
$^a$ Additional data point for $f=9.6$--20.1 Hz is also present \\
$^b$ Data for the QPO harmonic, $f=4$--5.5 Hz are also used \\
\label{tab:ffreqs}
\end{table*}

\subsection{XTE~J1550-564}

XTE~J1550-564 have been studied by many authors (e.g.\ Sobczak et al.\  
2000; Wilson \& Done 2001). We chose four observations from the
1998 outburst (Cui et al.\ 1999; Kubota \& Makishima 2004; Kubota \& Done 
2004) and one observation from the 2000 outburst (Tomsick, Corbel \& 
Kaaret 2001).

During our first observation (obs.\ 8 in Cui et al.\ 1999) the 
source already switched to a soft state. The spectrum cannot be described 
by a two-component continuum model (either a disc blackbody and a Comptonization,
or a two Comptonization model). A three component model provides a good fit: 
disc blackbody of $\Tbb \approx 0.6$ keV and its Comptonization on cool 
($kT_{\rm e} \approx 8$ keV) and hot ($kT_{\rm e} = 150$ keV) electrons. 
The data require a somewhat broadened, highly ionized 
($\xi \sim 10^3$) reflection ($\Or \sim 0.4$) and an additional narrow iron 
line at 6.4 keV.

Our next three observations are representative to the ,,anomalous'' very 
high state (Kubota \& Makishima 2004; Kubota \& Done 2004), which the source 
entered after being in the standard very high 
state during the 1998/1999 outburst. A detailed timing analysis of these 
observations was presented in, e.g., Remillard et al.\ (2002). We 
find a very good description of the (III) and (IV)  datasets 
with a model consisting of two 
Comptonizations. Additional disc blackbody component is required only in the 
first anomalous state observation (dataset II). 
The electron plasma temperatures are  $k T_{\rm e} = 5$--15 for the soft
Comptonization and it was fixed at 150 keV for the hard Comptonization.
The temperature of seed photons drops from $kT_{\rm bb} = 
0.9$ to 0.5 keV. The data require a reflection component whose strength 
also decreases (from $\Or \approx 2$ to 0.2). The reflecting medium is highly 
ionized ($\xi = 1300$--4000), and the data required also some relativistic 
smearing.  The first and third datasets need a narrow Gaussian line at 
6.4 keV.

Our last observation was taken during the 2000 outburst, when the source was 
transiting to a hard state. This is observation 3 in Kalemci et al.\ (2001) 
(timing analysis) and Tomsick et al.\ (2001) (spectral 
analysis). We found that the data can be very well described with a weak 
disc blackbody ($kT_{\rm bb} \sim 0.9$ keV) and thermal Comptonization 
($kT_{\rm e} = 150$ keV). The photon index of the continuum is very hard, 
$\Gamma \sim 1.7$. The data require a weak reflection component ($R \approx
0.1$) from an ionized medium ($\xi \approx 1700$), with significant smearing,
corresponding to relativistic smearing with inner disc radius of 
$8\,R_{\rm g}$). The residua at 6.4 keV are fit with 
additional narrow gaussian line.

\subsection{4U~1630-47}

We chose two observations of 4U~1630-47 from the beginning of the
1998 outburst, when the source was still in a low/hard state. A
detailed timing analysis of the outburst was performed in
Dieters et al.\ (2000). In particular, a complex behaviour of
QPO features during the transition was analysed. According to Dieters et
al.\ (2000) classification, based on the power density properties,
our observations are an example of type A. We find that the energy
spectra are well described by a model composed of a weak 
disc blackbody component with a relatively high soft photons temperature,
$kT_{\rm bb} \sim 1$, and its Comptonization on hot electrons.
The continuum slope is $\Gamma \sim 1.9$--2, 
intermediate between the hard slope of the last dataset of XTE~J1550-564
and the soft state slopes.  The data also require a weak reflection component 
($\Or \approx 0.1$) from an ionized matter ($\xi = 2000$--5000).

\section{QPO data analysis}

\begin{figure*}
\centering
\includegraphics[width=16cm,bb=18 360 592 720,clip]{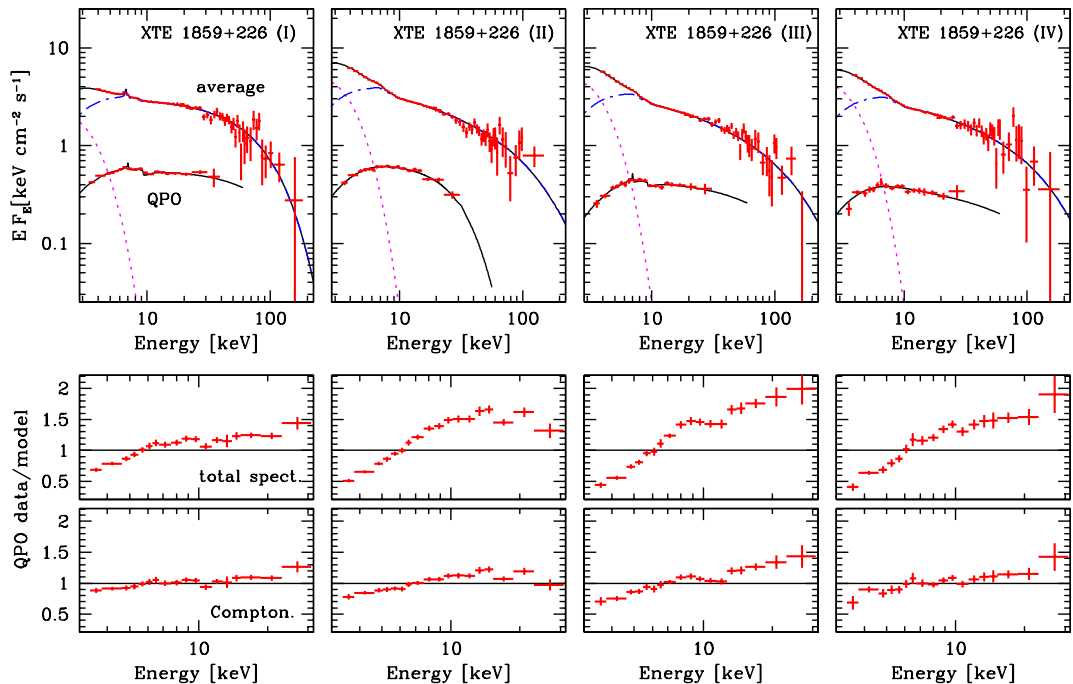}
\caption{A comparison of the QPO and mean energy spectra in the soft spectral
   state of XTE~J1859+226. {\it Upper panels\/}: time averaged and
   QPO energy spectra, unfolded using models described in 
   Sec.~\ref{sec:datasel} and Sec.~\ref{sec:reflection}, respectively. 
  Solid curves show the total spectra, the dotted curve (magenta online)
  is the disc blackbody component, while the dot-dashed curve (blue online) 
  is the Comptonized component.
 {\it Lower panels\/} show
   ratios of the QPO data to various model components:
   the total time averaged continuum (but without the reprocessed component), 
 and to the time averaged 
  Comptonization continuum only (i.e.\ the disc and reprocessed components 
  removed). The QPO spectra are {\em harder\/} than the time averaged
  spectra in this soft state, and they are closer to pure Comptonization
continua than to the total continua (i.e.\ the disc component does {\em not\/}
seem to contribute to QPO). The reprocessed component is present in two
QPO spectra (I and III; see Table~\ref{tab:qpo_meansp.refl}).}
\label{fig:spect_ss1}
\end{figure*}

\subsection{Light curves and power density spectra}

We used binned and event mode PCA data for the high time resolution spectral
analysis. PCA configurations used for each dataset are given in 
Table~\ref{tab:ffreqs}.
We generated light curves in each energy channel using standard 
tools from the {\sc ftools} package. We used the binned mode data in the 
range 3--13 keV and the event mode data above 13 keV.
We computed the power
density spectra in the Leahy normalization with the white noise
subtracted in {\sc powspec}. Then, we corrected each PDS for background and
renormalized to the Miyamoto normalization (i.e., (r.m.s./mean)$^2$), 
by multiplying by $\frac{S+B}{S^2}$, where $S$ and $B$ are source and 
background count rates, respectively (Berger \& van der Klis 1994).

\subsection{QPO r.m.s.\ spectra}
\label{sec:ffspectra}

We modeled the PDS either as a broken power law continuum with a Lorentz 
QPO profile, or as a sum of Lorentz 
profiles describing both the continuum and the QPO (e.g.\ Nowak 2000). 
These are 
phenomenological descriptions, however, since a Lorentz function is a 
Fourier transform of a damped harmonic oscillator, it can be assumed (in 
particular in the latter case) that each Lorentzian in the PDS is a 
signature of a quasi-periodic process with a different degree of 
coherence. The QPO energy spectrum was then obtained by integrating the 
QPO Lorentzians over $f$ for different energies.

We constructed response matrices for binned and event mode data using {\tt 
pcarsp} (the background correction was performed at the stage of 
constructing PDS) and we fit the {\it ff}-spectra in {\sc xspec} in order 
to quantify the spectral trends. We used data in a relatively wide energy 
range, $\sim 3$--40 keV, but since the PCA data become background 
dominated above 20--30 keV, any residua and features at 20--40 keV should 
be treated and interpreted with caution.


\section{Results}

Here we  compare the energy dependent QPO r.m.s.\ spectra
to the time averaged spectra. The latter were described in earlier
sections on individual objects. The results are presented in the form
of ratios of QPO data to the model components of the time averaged
spectra, when the model normalization was adjusted to give smallest
possible $\chi^2$. These will be useful when we discuss the data in the context
of theoretical models in the next Section. 
We will also perform formal model fits
to some of the QPO spectra in order to more quantitatively determine
their shape relative to the time averaged spectra.

Our sample contain 14 observations of four objects. Of these 11 observations
were performed in soft spectral states while 3 in the hard state.

\begin{figure*}
\centering
\includegraphics[width=16cm,bb=18 305 592 720,clip]{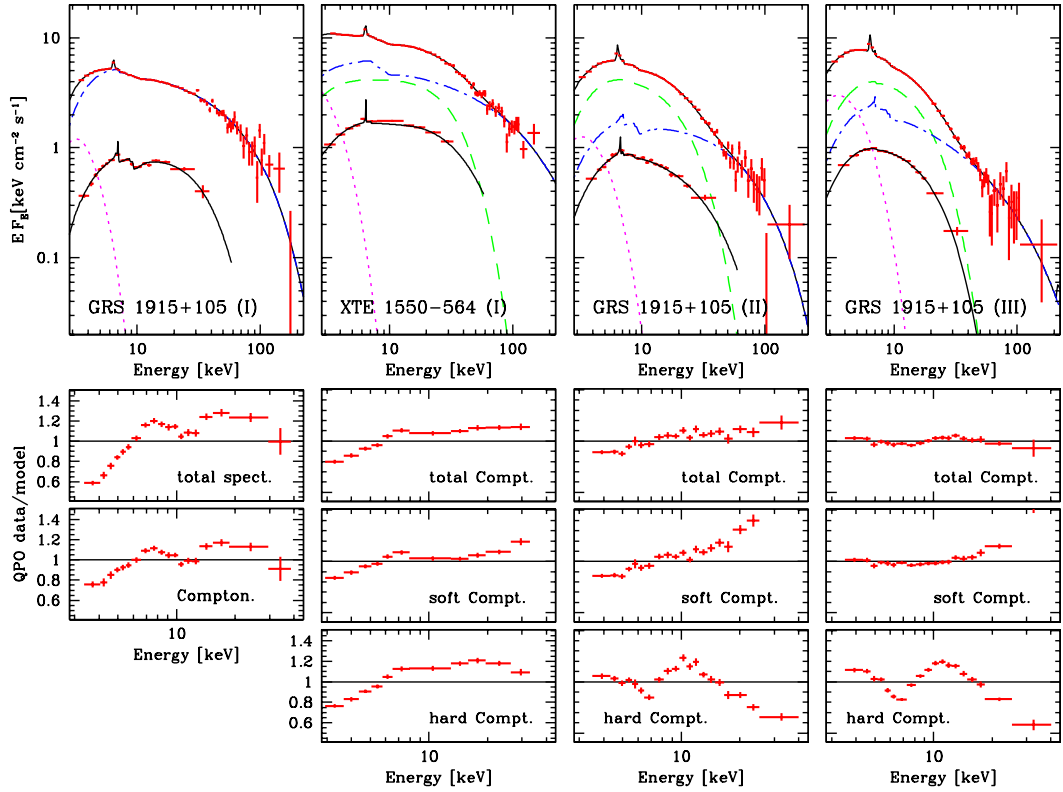}
\caption{A comparison of the QPO and mean energy spectra in the soft spectral
   states of GRS~1915+105 and XTE~J1550-564. Panels are described in details
   in Fig.~\ref{fig:spect_ss1}. The dashed (green online) curve is the soft
 Comptonization component. Ratios of the QPO data to: total continuum 
  spectrum, total (two-component) Comptonization, the soft Comptonization and
 the hard Comptonization, are plotted as labelled.}
\label{fig:spect_ss2}
\end{figure*}

\subsection{Soft state spectra}

In the soft state of XTE~J1859+226 (Fig~\ref{fig:spect_ss1}) the QPO spectra 
are more similar to the Comptonized component than to the total spectra.
This means that the disc blackbody component is not present in the QPO
spectra, i.e.\ it does not participate in the oscillations. The QPO
spectra are harder than the  Comptonized component in the corresponding
time averaged spectrum. Ratios
for dataset (III) clearly indicate the presence of the reprocessed
component in the QPO spectrum, which will be discussed in more details
below.

Similarly, no disc blackbody component in the QPO spectra
is observed for GRS~1915+105 (Fig.~\ref{fig:spect_ss2}).
In all three spectra the QPO data show larger deviations
from the total model than from the (total) Comptonized component.
In datasets (II) and (III), where the time averaged spectrum requires two
Comptonized components, the QPO are most similar to the sum of the two
Comptonizations, rather than to either single one of them. In particular,
the QPO spectra seem to be much harder than the soft Comptonization.

\begin{figure*}
\centering
\includegraphics[width=16cm,bb=18 280 592 720,clip]{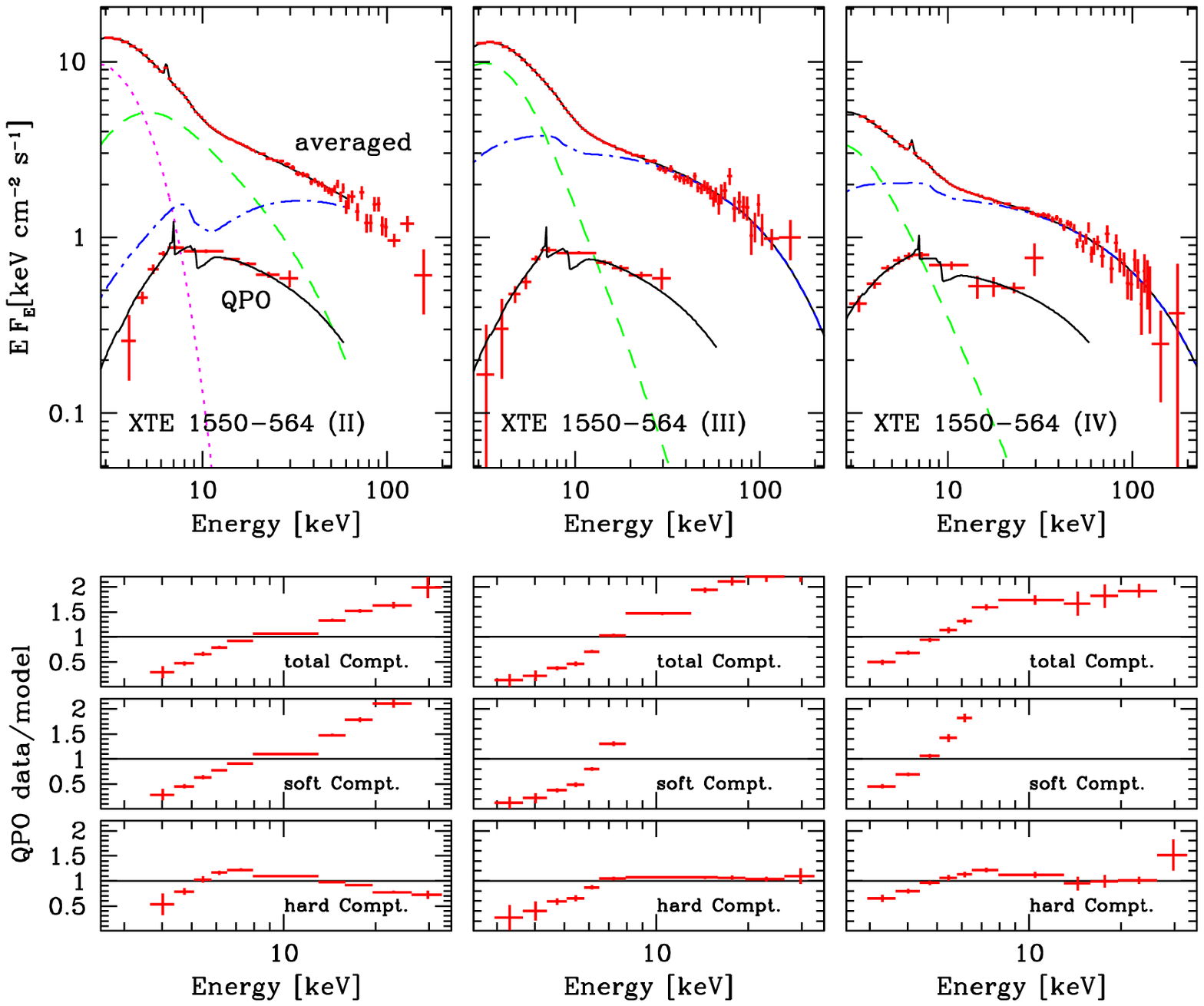}
\caption{A comparison of the QPO and mean energy spectra in the ``anomalous''
 very high state of XTE~J1550-564. Panels are described in details
 in Fig.~\ref{fig:spect_ss1}. Ratios of the QPO data to: total Comptonized
 continuum, soft Comptonized component and hard Comptonized component
 are plotted as labelled.}
\label{fig:qpo_meansp.1550}
\end{figure*}

We attempt to fit the QPO data with a modified time averaged Comptonization 
continuum. We first tie all the parameters of the QPO model, except 
for normalization, to the parameters of the time averaged spectrum. 
Fitting the GRS~1915+105 (I) dataset we also set the disc blackbody 
normalization to 0, so the only free parameter is the normalization of the
Comptonized component. The fit is very bad, $\chir=267/18$ dof. It improves
dramatically, when seed photon temperature, $T_0$, is free to adjust, 
$\chir=42.3/17$ dof. The fit improves further if the continuum slope
is left free, $\chir=35.8/16$. The resulting model has higher 
$T_0 \approx 0.95$ keV compared to the $\approx 0.65$ keV in the mean 
spectrum, and it is harder than the latter, $\Gamma\approx 2.18$ 
compared to $\approx 2.25$. The fit can be further improved if the
reflection amplitude is free to adjust, which will be discussed below.

For GRS~1915+105 datasets (II) and (III) there is a number of ways
the time averaged Comptonized components can be modified to fit the QPO
spectrum, since there are two Comptonized components in each spectrum 
(in addition to the disc blackbody). Letting $T_0$ (common to the
two components) to be free does provide a good fit, with the best fit value
again higher than that in the mean spectrum.
When the seed photon temperature is fixed, 
$T_{\rm 0,\ QPO} = T_{\rm 0, mean}$, obtaining
a fit of similarly good quality requires freeing at least three parameters:
the two spectral slopes and electron temperature of the soft Comptonization.

In the ``anomalous'' VHS of XTE~J1550-564 the QPO spectra are again harder
than the time averaged spectra (Fig.~\ref{fig:qpo_meansp.1550}). 
The disc blackbody component, even if present in the mean spectrum, 
is not present in the QPO spectra.
The QPO spectra slope above $\approx 7$ keV is similar to that of the hard
Comptonization slope in time averaged spectrum, but there is a strong
cutoff in QPO spectra below that energy, which formally can be described 
as a high value of $T_0$. We therefore test a model, where the QPO spectrum 
is equal to the hard Comptonization component in the mean
spectrum, but their common seed photon temperature is untied from the disc
blackbody temperature or soft Comptonization seed photon temperature,
$T_{\rm 0,QPO}=T_{\rm 0,hard} \ne T_{\rm 0,soft} = T_{\rm bb}$.
Such a model gives somewhat worse fits to the combined datasets than best 
models to time averaged data alone. Pairs of $\chir$ values  for 
the mean spectrum fit (from Section~\ref{sec:datasel}) and for the combined
mean spectrum and the QPO spectrum fits, are as follows:  
($65.8/69=0.95$, $90.8/77=1.18$), 
($51.3/71=0.71$, $65.6/81=0.81$) and 
($74.5/73=1.02$, $85.1/82=1.04$), for dataset (II), (III) and (IV), 
respectively. For datasets (III) and (IV) the increase of $\chi^2$ is
14.3 and 10.6 per 10 and 9 new dof, respectively. Thus including the QPO data 
does not significantly worsens these fits. However, when the QPO fits are 
examined as data/model ratios, they are approximately the same for all 
datasets. Its the larger error bars on datasets (III) and (IV) (compared to
dataset I) that make the fits acceptable. In particular the data show
pronounced low energy cutoffs below $\approx 6$ keV, which is not well 
modelled by the increased $T_0$ if the latter is also to fit the mean spectrum.
We thus conclude that the QPO data are
unlikely to be described simply by the same shape as the hard Comptonizing 
continuum. This implies that more complicated models, which include spectral
variability, must be employed to explain the QPO in these datasets.
We intend to explore such models in more details in a future paper.

\subsection{Hard state spectra}

\begin{figure*}
\centering
\includegraphics[width=16cm,bb=18 320 592 720,clip]{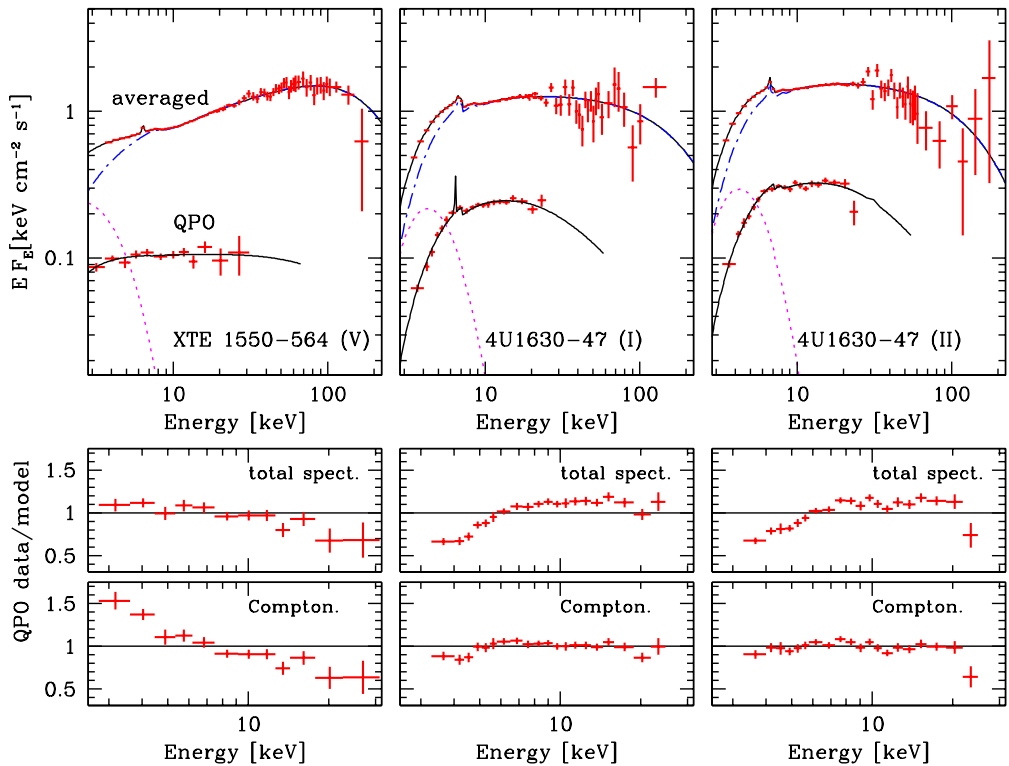}
\caption{A comparison of the QPO and mean energy spectra in the hard spectral
   states. Panels are described in details
   in Fig.~\ref{fig:spect_ss1}. Ratios of the QPO data to: total continuum 
  spectrum and the Comptonized continuum are plotted as labelled.}
\label{fig:qpo_meansp.hs}
\end{figure*}

In the hard spectral state of XTE~J1550-564 (dataset V; 
Fig.~\ref{fig:qpo_meansp.hs}),
the QPO spectrum is somewhat softer than the total spectrum. When the
model for the time averaged spectrum is fitted to the QPO data,
the best fit model does contain the disc blackbody component 
($\Tbb \approx 0.89$ keV). When the disc blackbody component is removed
and the seed photon temperature, $T_0$, is fixed at $\Tbb$, the fit is worse by
$\Delta\chi^2 = 7.2$ for one more dof ($\chir=13.1/10$ vs.\ $5.7/9$).
Now, if $T_0$ is free to adjust the fit improves to $\chir=5.7/9$, but
the new $T_0 = 0.54_{-0.54}^{+0.23}$ is {\em lower\/} than $\Tbb$. Thus,
contrary to the QPO spectra in soft states, here the QPO spectrum either 
contains the disc blackbody component, or the seed photon temperature
in QPO spectra is lower than that in time averaged spectra. In either case
the QPO spectrum slope is softer than the time averaged spectrum.

\begin{figure}
\centering
\includegraphics[height=7.5cm,angle=270,clip]{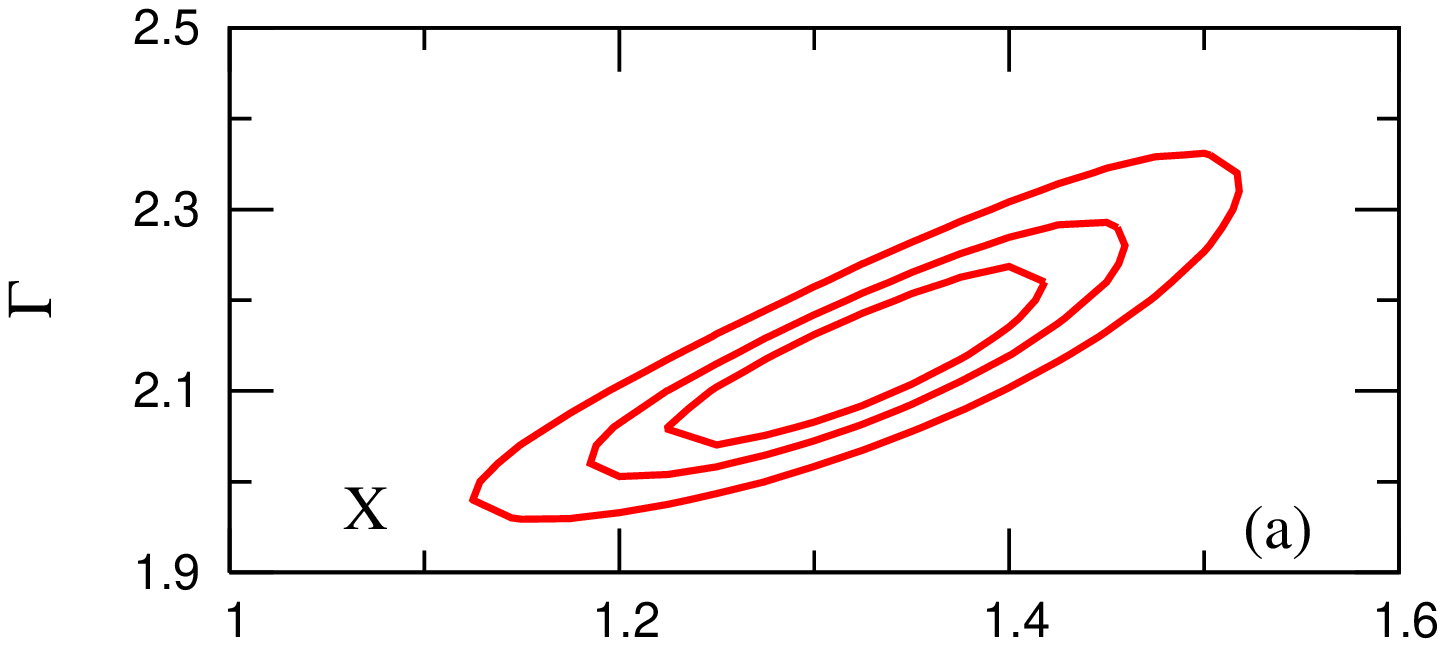}
\includegraphics[height=7.5cm,angle=270,clip]{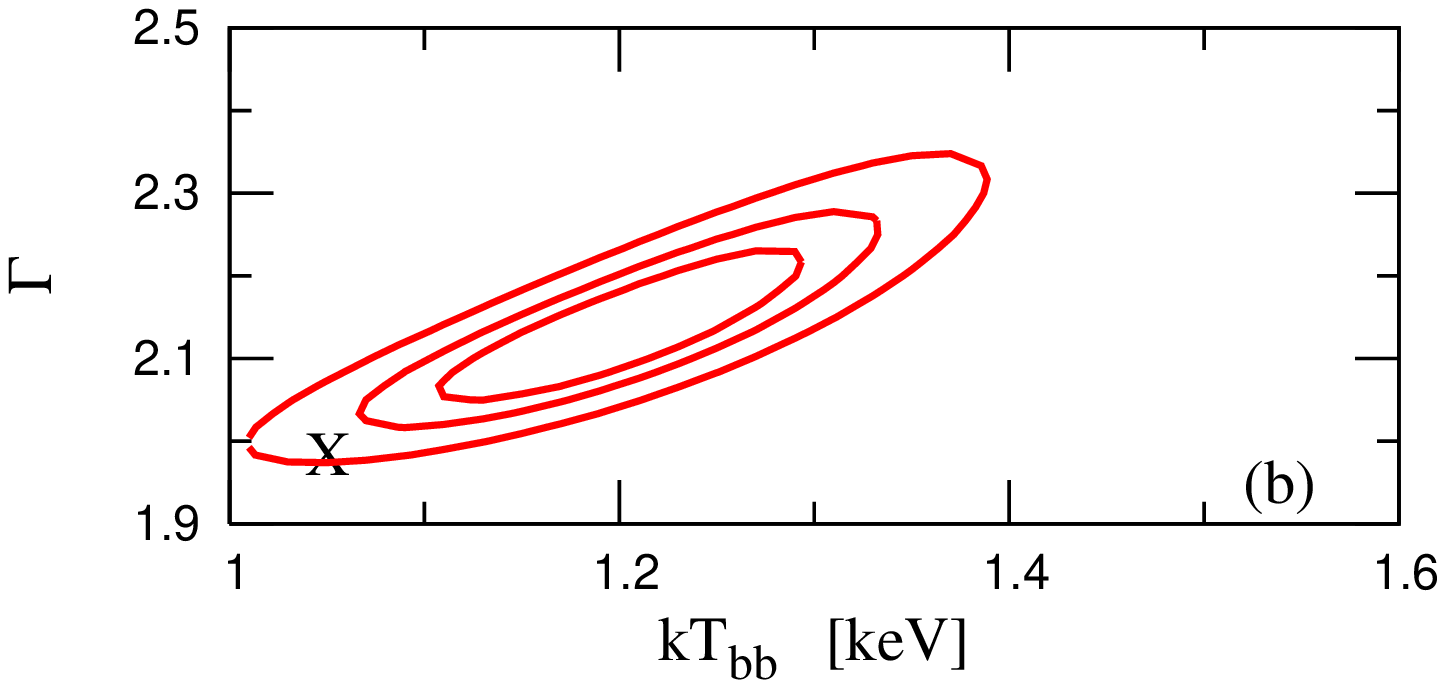}
\caption{Confidence contours ($\Delta\chi^2=2.3,\ 4.6\ {\rm and}\ 9.6$) in the 
$\Gamma$--$T_0$ plane for the QPO spectra from the observations of 4U~1630-47 in
the hard state. 'X' mark the best-fit parameters to the time averaged spectra.
The QPO spectra are somewhat softer and have higher $T_0$ than the time averaged 
spectra}
\label{fig:qpo_hard_cont}
\end{figure}

In the other two hard state observations, of 4U~1630-47, the QPO slope 
seems to be similar to the Comptonization slope in the time averaged
spectrum, but the residuals suggest somewhat higher $T_0$ 
(Fig.~\ref{fig:qpo_meansp.hs}). When the time averaged model is renormalized
($\Gamma$ and $T_0$ fixed) to the QPO data, the fits have $\chir=27.5/19$ and 
$\chir = 25.4/19$ for datasets (I) and (II), respectively. Allowing $\Gamma$
and $T_0$ to adjust we obtain best fits with $\chir=11.0/17$  and $\chir=17.5/17$,
and the confidence contours are plotted in Fig.~\ref{fig:qpo_hard_cont}.
The contours suggest that the QPO spectra are again
somewhat softer and indeed have higher $T_0$ than the time averaged spectra,
however the difference between the spectral slopes in not highly significant.
We note then that for 4U~1630-47 the QPO spectra are closer to the time 
averged spectra than those of XTE~J1550-564 (V). Since the time averaged
spectra 4U~1630-47 are softer than the spectra of XTE~J1550-564 (V), 
we might be observing here a transition from the hard state behaviour (as in 
XTE~J1550-564) to the soft state behaviour, where the QPO spectra are
harder than the time averaged spectra.

\subsection{The reprocessed component in the QPO spectra}
\label{sec:reflection}

We have checked if the reprocessed component is  present in the QPO spectra.
Firstly, we assume the time averaged continuum as a model for the QPO,
and let the reflection parameters, $\Or$, $\xi$ and $\Rin$ to be free.
Such models do not provide good fits, with only two cases when $\chir<1.5$
(Table~\ref{tab:qpo_meansp.refl}; columns {\it a} and {\it b}). Therefore,
although the reprocessed component is present in the model, it is difficult
to assess its statistical significance. In the two cases where the fits
are acceptable (XTE~J1859+226, datasets I and IV), the reflection is
significant.

Secondly, we fit the QPO data with the {\sc thComp} model allowing
all its parameters, $\Gamma$, $k T_{\rm e}$ and $T_0$ to adjust.
The continuum is thus described correctly, which allows for meaningful 
determination of the presence  and parameters of the reprocessed component.
We then add the reprocessed component, assuming first 
a cold matter ($\xi=0$), and then allowing $\xi$ to adjust as well.
In Table~\ref{tab:qpo_meansp.refl} we present the quality of the initial
model (i.e.\ no reflection) fits, and the $\Delta\chi^2$ values when the 
reprocessed components are added.  

Adding the reprocessed component improves the fits in 
most of the datasets, with ionized reprocessing giving better fits than
cold reprocessing. Reduction of $\chi^2$ values by 10--60 means that
the reprocessed component is highly significant.

Reflection amplitudes in the QPO spectra are generally rather poorly 
constrained. 
In the three anomalous very high state spectra of XTE~J1550-564 the best
fits are obtained for a pure reflected component. However, the ionization
parameter in the pure reflection models is rather high, $\xi \sim 10^5$,
which means that the Fe reprocessing features are not very prominent.
The 90 per cent lower limits to $R$ are rather low, nevertheless, 
reduction of $\chi^2$ by 11-22 means that the component is highly significant
in at least two of those spectra.

\section{Discussion}

We have applied techniques of Fourier spectroscopy to the analysis of X-ray 
spectra of a number of black hole X-ray binaries showing QPO in their PDS and 
we have analysed the QPO 
r.m.s.\ spectra. We find that in general the quasi-periodically variable part of 
X-ray 
spectra are consistent with thermal Comptonization and they do not require 
the disc thermal component for their description. This extends previous 
studies of Churazov et al.\ (2001) (for soft state of Cyg X-1), 
and it argues for the origin of the amplitude of the variability
(including the QPO) in the hot plasma rather than in the cold disc.

\begin{table*}
\caption{Test for the presence of the reprocessed component in the QPO spectra}
\centering
\begin{tabular}{l c c c c c c c }
\hline
Object & $\chi^{2,a}/{\rm dof}$ & $\Or^{b}$ &  
 $\chi^{2,c}/{\rm dof}$ &  $\Delta\chi^{2,d}$ & $\Or^{e}$ &  
  $\Delta\chi^{2,f}$ & $\Or^{g}$ \\ 
\hline
XTE~J1859+226 (I)   & 18.7/16 & $0.47 \pm 0.17$ & 
                      27.0/16 & -3.7  & 0.3  & -18.2 & $>0.8^{f}$ \\
XTE~J1859+226 (II)  & 76.5/16 & 0.72 &
                      12.4/14 &  0.0  & $0^{+0.3}$  & 0.0  & $0^{+0.4}$ \\
XTE~J1859+226 (III) &  24.9/14  & 2.2 &
                      16.5/15 & -2.7  & 0.50 & -11.0 & $> 0.2$ \\
XTE~J1859+226 (IV)  &  17.0/15  & $0.90^{+0.35}_{-0.30}$ &
                      9.86/15 & -0.85 & 0.42 & -0.9 &  $0.2^{+3.5}_{-0.2}$ \\
\hline
XTE~J1550-564 (I)   &  17.5/7  & 1.7  &
                       7.2/7   & -5.2  & $0.62^{+0.63}_{-0.47}$ & --    & --  \\
XTE~J1550-564 (II)  &   65.1/6 & 0.6 &
                      30.8/7  & -15.3 & 2.1  & -22.1 & $>0.21^{h}$ \\ 
XTE~J1550-564 (III) &   84.6/8 &  1.3 &
                       15.7/8  & -4.9  & 1.3  & -11.7 & $>0.16^{h}$ \\
XTE~J1550-564 (IV)  &  18.0/6  &  3.0 &
                       10.9/8  & -2.5  & 1.2  & -4.8  & $>0.08^{h}$ \\
\hline
\hline
XTE~J1550-564 (V)  &   24.2/9 & 0.02 &
                      5.7/9  & --   & --  & --  & -- \\
\hline 
4U~1630-47 (I)    &  25.7/17  & 0.18 &
                      15.4/17 & -5.4 & $0.77^{+0.73}_{-0.58}$ & -- & -- \\
4U~1630-47 (II)   &  19.5/16  & $0.23^{+0.21}_{-0.13}$ &
                       21.4/17 & -1.9 & 0.40 & -6.1 &  $1.00^{+2.5}_{-0.95}$ \\
\hline
GRS~1915+105 (I)  & 45.4/15 & 1.8 &
                    76.4/16 & -30. & 1.7  & -65.9 & $> 0.8$ \\
GRS~1915+105 (II) & 31.0/16 & 1.7 &
                     27.9/16 & -7.6 & 0.97 & -10.8 & $0.21^{+0.46}_{-0.12}$ \\
GRS~1915+105 (III)&  --  & -- &
                       7.9/15 & --   & --   & -- & -- \\
\hline       
\end{tabular}\\
$^a$ $\chi^2$/dof from fits of the time averaged spectrum model to the QPO 
spectra, allowing {\em only} the parameters of the reprocessed  component 
and overall normalization to be free \\
$^b$ Best fit value of $\Or$  from fits $^a$ \\
$^c$ $\chi^2$/dof from fits of the {\tt thComp} model to the QPO spectra
(all parameters adjustable), 
{\em without\/} the reprocessed  component \\
$^d$ $\Delta\chi^2$ when cold ($\xi=0$) reprocessed component is added,
with no relativistic smearing (i.e., one additional parameter)\\  
$^e$ Best fit value of reflection amplitude (or 95 per cent upper limit, 
 $\Delta\chi^2=2.71$, when $\Delta\chi^2=0$ in column $^d$) \\
$^f$ $\Delta\chi^2$ (relative to fits $^c$) when ionized reprocessed 
 component is added (2 additional parameters) \\
$^g$ Best fit value of $\Or$  from fits $^f$ \\
$^h$ Best fit obtained for a pure reflected component ($\Or=+\infty$);  
 95 per cent lower limit shown \\
$^i$ the disc blackbody component was included in the QPO model
\label{tab:qpo_meansp.refl}
\end{table*}

Comparing the time averaged and the QPO spectra we explicitly find that 
only one component, a thermal Comptonization (with its Compton reflection)
is needed to describe the 
latter. However, the QPO spectra are in most cases significantly different 
than the Comptonized component in the time averaged spectra. 
Qualitatively, we find an interesting anti-correlation: in soft spectral 
states the QPO spectra are harder than the mean spectra, while they are 
softer than the mean spectra in hard spectral states. Together with two 
intermediate cases this forms a continuous sequence. Quantitatively, 
the differences between the spectra can be described by a difference in 
spectral indices or seed photons temperatures, or both, with 
the statistical quality of the 
various descriptions being approximately the same.

The QPO spectra do contain the reprocessed component. In most datasets
their presence is highly statistically significant (99 per cent or better),
although the reflection amplitude is rather poorly constrained. 
Because of that, it is not possible to determine if the amplitude is
different than that in the time averaged spectrum.

In Paper I we introduced a phenomenological model of QPO, where the
background radial propagation model of broad band variability (Kotov et 
al.\ 2001, \.{Z}ycki 2003) is supplemented with quasi-periodic modulation 
of one or more of 
the parameters of a Comptonization spectrum. The main parameters include 
the plasma heating rate, $\lheat$, the cooling rate by soft disc photons,
$\lsoft$ (here $l \equiv (L/R)\sigma_{\rm T}/(m c^3)$ is the compactness
parameters), and the amplitude 
of the reflected component (describing the coupling between the heating 
and cooling). Our motivation for such a model is that irrespectively of 
the physical mechanism behind the QPO, the modulation of the emitted 
X-rays can only be produced by a modulation of one or more of the 
parameters actually 
determining the hard X-ray spectrum. We can now discuss our observational 
results in the light of that model.

QPO spectra harder than the mean spectra can be obtained if the QPO are 
produced by modulation of the heating rate while the cooling rate does not 
respond fully to $\lheat$ modulation. This produces 
spectral pivoting around the low energy end of the spectrum, and, in 
consequence, r.m.s.\ variability increasing with energy. Such a model 
might then be applicable to the soft spectral states discussed in this paper.
The model predicts a local maximum of the EW of the Fe \Ka line at the QPO
frequency (see fig.~3 in Paper I), but the quality of our data is not
sufficient to test this prediction.

QPO spectra softer than the mean spectra can be produced by modulations 
driven by the cold disc. One possibility here is a modulation of the cooling 
rate.
This results in the spectra pivoting around an energy point intermediate 
between  the low and high energy ends of the spectrum (see also Zdziarski et 
al.\ 2002). In the limited energy band, e.g.\ that of {\it RXTE}/PCA, it may 
correspond to the r.m.s.\ spectra decreasing with energy. 

The cooling rate may be modulated with or without modulation of the seed 
photons temperature. 
The difference between the two is that the QPO spectrum is monotonic in energy
in the former case but it has a very deep minimum (at the pivot energy) 
in the latter case (see figs.~4 and 5 in Paper I). 
Additionally, QPO harmonics appear when
$\Tbb$ is modulated, since the modulation of $\lsoft$ is then not sinusoidal
(even if modulation of $\Tbb$ is). However, the harmonics are limited to
the soft disc component only, unlike in some of our data, for example, 
XTE~J1550-564~(I).

The other possibility of producing a soft QPO spectrum is to modulate the 
amplitude of the reflected component, $\Or$, assuming that it also describes 
the 
feedback between illuminating hard X-rays and the re-emitted soft disc photons 
cooling the hot plasma. Because of the coupling, this again generates 
modulations of $\lsoft$ and the characteristic spectral pivoting.
The most characteristic feature of this model is the strong
Fe \Ka line (and the entire reprocessed component) at $\fQPO$, directly
resulting  from modulations of $\Or$. Contrary to that, the model with 
$\lsoft$ modulation gives either generally weak Fe line, or at least a 
minimum at $\fQPO$. The presence of the reprocessed component (including
the Fe \Ka line) in our QPO spectra makes modulation of $\Or$ an
interesting possibility (see also Miller \& Homan 2005). 

We note that models from Paper I involving modulation of $\lsoft$ predict 
very strong soft component in the QPO spectra, which is not observed.
It is rather hard to envision a geometry where the modulated soft flux
would enter the hot plasma, but would not be directed towards an observer.
Modulation of $\Or$ also produces some soft component in the QPO spectra,
but it appears less prominent than the one from $\lsoft$ modulation models.
Clearly, further development of the models is necessary to fully describe 
the data.

Another potentially useful diagnostics is the coherence function. While
variations in different energy bands are perfectly coherent when $\Or$ is 
modulated, the coherence function shows a complex behaviour, with a number
of minima, when $\lsoft$ and/or $\Tbb$ are modulated. However, distinguishing 
between the models based on such predicted observables as the EW of the 
Fe \Ka line, the coherence function or the hard X-ray time lags 
would require better data than currently available.

It needs to be emphasized that majority of physical models of QPO 
envision the QPO as driven by some kind of disc oscillations.
However, our interpretation of the energy dependencies of low-$f$ QPO, at
least in soft spectral states, would rather point out to modulations
of the heating rate in the hot plasma. If so, then those models must find 
a way of transferring the disc oscillations to the hot plasma with 100 per cent
efficiency, i.e.\ without affecting the disc emission.
On the other hand, most of the models were formulated in 
the context of the high frequency QPO, where the energy 
dependencies might be different than those for the low-$f$ QPO.
The only work where a physical model of the low-$f$ QPO was 
constructed is that by Giannios \& Spruit (2004). They consider oscillations 
of a hot ion-supported accretion flow coupled to a outer cold disc. The
coupling is realized by a number of channels: hard X-rays illuminating the
cold disc, hot protons heating the cold disc, the soft disc photons cooling 
the hot flow. Spectral variability predicted by that model corresponds
to our case of modulating the cooling rate, that is, the QPO spectra are
softer than the time averaged spectra. This would then be consistent with
observed QPO behaviour in the hard state. The overall geometry of a hot inner
flow and an outer cold disc fits that state too.

Considering that the usual broad band noise X-ray variability is driven
by instabilities and flares in the hot plasma, it is obvious that an important
piece of physical understanding is still missing.

\section{Conclusions}

\begin{itemize}
 \item The QPO energy spectra are harder than the time average spectra in
soft spectral states, but they are softer than the time averaged spectra
in the hard state. The QPO spectra are similar in slope to the time
averaged spectra, when the latter are intermediate in slope between hard and soft
($\Gamma \approx 2$).
 \item The disc component is absent in the QPO spectra.
 \item Comparison of the observational data with simple models of spectral
variability suggests that instabilities in the hot plasma drive the 
low-$f$ QPO in the soft state, while cold disc oscillations drive
the QPO in the hard state.
\end{itemize}

\section*{Acknowledgments}
This work was supported in part by  Polish Committee of Scientific Research
through grants number 1P03D01626 and 2P03D01225.


{}

\label{lastpage}



\begin{thebibliography}{}

\bibitem[]{}
  Belloni T.,  van der Klis M., Lewin W.~H.~G., van Paradijs J., Dotani T.,
   Mitsuda K., Miyamoto S., 1997, A\&A, 322, 857

\bibitem[]{}
  Belloni T., Klein-Wolt M., M{\' e}ndez M., van der Klis M., van Paradijs J.,
   2000, A\&A, 355, 271

\bibitem[]{}
   Berger M., van  der Klis M., 1994, A\&A, 292, 175

\bibitem[]{}
  Casella P., Belloni T., Homan J., Stella L., 2004, A\&A, 426, 587

\bibitem[]{}
  Churazov E., Gilfanov M., Revnivtsev M., 2001, MNRAS, 321, 759


\bibitem[]{}
  Cui W., Zhang S.~N., Chen W., Morgan E.~H., 1999, ApJ, 512, L43

\bibitem[]{} 
  Dieters S.~W. et al., 2000, ApJ, 538, 307

\bibitem[]{} 
  Done C., Wardzi\'{n}ski G., Gierli\'{n}ski M., 2004, MNRAS, 349, 393

\bibitem[]{} 
  Done C., \.{Z}ycki P.~T., Smith D.~A., 2002, MNRAS, 331, 453

\bibitem[]{} 
  Giannios D. \& Spruit H. C, 2004, A\&A, 427, 251

\bibitem[]{} 
 Gierli\'{n}ski M., Zdziarski A. A., 2005, MNRAS, 363, 1349

\bibitem[]{} 
  Gilfanov M., Revnivtsev M., Molkov S., 2003, A\&A, 410, 217

\bibitem[]{} 
  Kalemci E., Tomsick J.~A., Rothschild R.~E., Pottschmidt K., Kaaret P., 
  2001, ApJ, 563, 239

\bibitem[]{}
  Kotov O., Churazov E., Gilfanov M., 2001, MNRAS, 327, 799

\bibitem[]{}
  Kubota A., Done C., 2004, MNRAS, 353, 980

\bibitem[]{} 
  Kubota A., Makishima K., 2004, ApJ, 601,  428

\bibitem[]{} 
  Maccarone T. J., Coppi P. S., 2003, MNRAS, 338,189

\bibitem[]{} 
  Maccarone T. J., Coppi P. S., Poutanen J., 2000, ApJ, 537, L107

\bibitem[]{} 
  Miller J. M., Homan J., 2005, ApJ, 618, L107

\bibitem[]{} 
  Nowak M. A., 2000, MNRAS, 318, 361

\bibitem[]{} 
  Negoro H., Kitamoto S., Mineshige S., 2001, ApJ, 554, 528

\bibitem[]{} 
  Poutanen J., 2001,  AdSpR, 28, 267 (astro-ph/0102325)

\bibitem[]{} 
  Poutanen J., Fabian A. C., 1999, MNRAS, 306, L31

\bibitem[]{} 
  Psaltis D., Belloni T., van der Klis M., 1999, ApJ, 520, 262

\bibitem[]{}
   Psaltis D., 2001, AdSpR, 28, 481 (astro-ph/0012251)

\bibitem[]{} 
  Rao A.~R., Naik S., Vadawale S.~V., Chakrabarti S.~K., 2000, A\&A, 360, L25

\bibitem[]{} 
  Remillard R.~A., Sobczak G.~J., Muno M.~P., McClintock J.~E., 
  2002, ApJ, 564, 962

\bibitem[]{} 
  Revnivtsev M., Gilfanov M., 2005, AN, 326, 812

\bibitem[]{} 
  Revnivtsev M., Gilfanov M., Churazov E., 1999, A\&A, 347, L23

\bibitem[]{} 
  Revnivtsev M., Gilfanov M., Churazov E., 2001, A\&A, 380, 520

\bibitem[]{}
  Rodriguez J., Corbel S., Tomsick J.~A., 2003, ApJ, 595, 1032

\bibitem[]{}
   Rodriguez J., Corbel S., Hannikainen D.~C., Belloni T., Paizis, A., 
  Vilhu O., 2004, ApJ, 615, 416

\bibitem[]{} 
  Rutledge R.~E. et al., 1999, ApJS, 124, 265

\bibitem[]{}
   Sobczak G.~J., McClintock J.~E., Remillard R.~A., Cui W., Levine A.~M., 
  Morgan E.~H.,  Orosz J.~A.,  Bailyn   C.~D.\ 2000, ApJ, 544, 993

\bibitem[]{} 
  Sobolewska M.~A., \.{Z}ycki P.~T., 2003, A\&A, 400, 553

\bibitem[]{}
   Tomsick J.~A.,  Kaaret P., 2001, ApJ, 548, 401

\bibitem[]{}
   Tomsick J.~A., Corbel S., Kaaret P., 2001, ApJ, 563, 229

\bibitem[]{}
   van der Klis M., 2006, in Lewin W. H. G., van der Klis M., eds, 
   Compact Stellar X-ray sources, Cambridge Univ. Press, Cambridge, 
   p.~39 (astro-ph/0410551) 

\bibitem[]{} 
   Wijnands R., 2001, Adv.\ Sp.\ Res., 28, 469

\bibitem[]{} 
  Wilson C.~D., Done C., 2001, MNRAS, 325, 167

\bibitem[]{}
   Zdziarski A.~A., Johnson W. N., Magdziarz P., 1996, MNRAS, 283, 193

\bibitem[]{}
   Zdziarski A.~A., Lubi\'{n}ski P., Smith D.~A., 1999, MNRAS, 303, L11

\bibitem[]{}
   Zdziarski A.~A., Poutanen J., Paciesas W. S., Wen L., 2002, ApJ, 578, 357

\bibitem[]{} 
  \.{Z}ycki P.~T., 2002, MNRAS, 333, 800

\bibitem[]{} 
  \.{Z}ycki P.~T.\ 2003, MNRAS, 340, 639

\bibitem[]{}
  \.{Z}ycki P.~T., Sobolewska M.~A.\ 2005, MNRAS, 364, 891

\bibitem[]{}
  \.{Z}ycki P.~T., Done C., Smith D. A., 1999, 305, 231

\end{thebibliography}
\end{document}